\documentclass[12pt]{iopart}
\usepackage[dvips]{graphicx}
\usepackage{epsf}

\def\ltsima{$\; \buildrel < \over \sim \;$}
\def\simlt{\lower.5ex\hbox{\ltsima}}

\begin{document}

\title[Gravitational waves, inflation and the cosmic microwave background]
{Gravitational waves, inflation and the cosmic microwave background: towards
testing the slow-roll paradigm}

\author{Carlo Ungarelli\dag, Pier Stefano
Corasaniti\ddag, R A Mercer\dag \, and Alberto Vecchio\dag}

\address{\dag \, School of Physics and Astronomy, The University of
Birmingham, Edgbaston, Birmingham, B15~2TT, UK}

\address{\ddag \, ISCAP, Columbia University, New York, NY~10027, USA}

\ead{ungarel@star.sr.bham.ac.uk}

\begin{abstract}
One of the fundamental and yet untested predictions of inflationary
models is the generation of a very weak cosmic background of gravitational 
radiation. We investigate the sensitivity required for a space-based 
gravitational wave laser interferometer with peak sensitivity at $\sim 1$ Hz 
to observe such signal as a function of the model parameters and compare it 
with indirect limits that can be set
with data from present and future cosmic microwave background missions.
We concentrate on signals predicted by slow-roll single field inflationary 
models and instrumental configurations such as those proposed for the
LISA follow-on mission: Big Bang Observer. 
\end{abstract}

%Uncomment for PACS numbers title message
%\pacs{00.00, 20.00, 42.10}

% Uncomment for Submitted to journal title message
%\submitto{\CQG}

\section{Introduction}
\label{sec:intro}

The paradigm of \emph{inflation}~\cite{Guth81,Linde82,AS82} emerged in
the early Eighties as a way of resolving a number of outstanding puzzles
in cosmology, by postulating that the Universe underwent a phase of
accelerated expansion. Inflationary models predict that the Universe is
spatially flat, and that the quantum zero-point fluctuations of the
space-time metric produce a nearly scale invariant spectrum of density
perturbations that are responsible for the formation of cosmic structures
and the generation of a primordial cosmic gravitational wave background
(CGWB). Observations of the Cosmic Microwave Background (CMB), most
recently with WMAP, have provided a confirmation of the first two
predictions~\cite{WMAP}; the generation of primordial gravitational waves
is still to be verified. This test is important for both cosmology and
fundamental physics. In fact, the actual detailed implementation of an
inflationary model requires the introduction of additional fields that
are not part of the already  experimentally well tested standard model
of particle physics and may produce effects at energy scales well beyond
those probed by particle physics experiments. The observation
of a CGWB either directly, with gravitational wave instruments, or
indirectly, via the effect on the CMB provides a unique way of measuring
the physical parameters of the models and an opportunity for testing new
ideas in fundamental physics and cosmology.

Inflation predicts a quasi-scale invariant CGWB between $\sim 10^{-16}$~Hz
and $\sim $1~GHz whose spectrum $h_0^2 \, \Omega_\mathrm{gw}(f)$ (the
fractional energy density in gravitational waves, normalised to the
critical density, per unit logarithmic frequency interval) does not
exceed $10^{-15}$ at any one frequency~\cite{Turner97}. Third generation
ground-based km-scale laser interferometers are expected to achieve a
sensitivity $h_0^2 \, \Omega_\mathrm{gw}(f) \sim 10^{-11}$ 
in the frequency range
$\approx 10$~Hz - a few $\times 100$~Hz (cf~\cite{CT02} for a
recent review). As the characteristic amplitude $h_\mathrm{c}$ on a
bandwidth $\Delta f$ produced by a stochastic background is

\begin{equation}
h_\mathrm{c}(f) \approx 4 \times 10^{-30} \, \left( \frac{h_0^2 \,
\Omega_\mathrm{gw}}{10^{-16}} \right)^{1/2} \left( \frac{f}{1 \,
\mathrm{Hz}}\right)^{-3/2} \left( \frac{\Delta f}{10^{-7} \, \mathrm{Hz}}
\right)^{1/2} \,,
\end{equation}
there is an obvious advantage in observing at lower frequencies.
Unfortunately, the Laser Interferometer Space Antenna
(LISA)~\cite{LISA_ppa} will not offer an opportunity to improve (much)
beyond the sensitivity of ground-based detectors because of the
instrument's limitations -- only one interferometer, preventing
cross-correlation experiments -- and the intensity of astrophysical
foregrounds in the mHz frequency band~\cite{HBW90,FP03,UV01}, where LISA
achieves optimal sensitivity. It is currently accepted that a LISA
follow-up mission aimed at the lowest possible frequency band not
compromised by astrophysical foregrounds, $0.1\,\mathrm{Hz} - 1\,\mathrm{Hz}$
represents the best opportunity to directly study inflation. As a result of this, a
new mission concept has recently emerged: the Big-Bang-Observer (BBO),
which is presently being investigated by NASA~\cite{bbo}. This consists
of a constellation of four interferometers in a Heliocentric orbit at 1
AU from the Sun. By making the arm length of the BBO interferometers
$\approx 100$ shorter than those of LISA, the centre of the
observational window is shifted to several $\times 0.1$ Hz; improved technology for
lasers, optics and drag-free systems will allow to achieve a sensitivity
$h_0^2 \, \Omega_\mathrm{gw}(f) \simlt 10^{-16}$. A similar mission, although
consisting of only one interferometer, has been proposed in Japan:
DECIGO~\cite{decigo}.

Gravitational waves produced during inflation will also have an indirect
effect on the structure of the cosmic microwave background (CMB) by affecting most 
importantly its polarisation~\cite{SZ97}. 
The investigation of the signature of GWs has been
one of the drivers in the design of Planck~\cite{planck}, an ESA mission
currently scheduled for launch in 2007; moreover vigorous efforts are
underway to design and develop more ambitious instruments, such as
CMBPol~\cite{cmbpol}, in order to carry out highly sensitive searches. 

The programme to test the prediction of the generation of a gravitational
wave stochastic background during inflation relies therefore on substantial sensitivity 
improvements for mission either in the gravitational wave or microwave band
(cf {\em e.g.}~\cite{Cooray05,SigCoo05}). 
In this paper we investigate how the direct observation of primordial 
gravitational waves by BBO can constrain the parameter space of inflationary 
models and what are the implications for the design of a mission. We also explore 
how such information compare with and complement those that can be gained with 
future CMB data. The paper is organised as follows: in Section~\ref{sec:model} we review
single-field slow-roll inflation, the spectrum $\Omega_\mathrm{gw}(f)$ of the cosmic
gravitational wave background  that is generated in this epoch 
and show that $\Omega_\mathrm{gw}(f)$ 
can be characterised by only two unknown parameters; in
Section~\ref{sec:results} we discuss the region of the parameter space
that can be probed by the Big-Bang-Observer mission, and how this
depends on different technological choices for the mission; we also
compare and contrast this results with what one might be able to 
achieve with future CMB observations, with missions such as Planck and CMBPol;
Section~\ref{sec:conclusions} contains our conclusions and pointers to
future work.

\section{Single-field slow roll inflation}
\label{sec:model}

In this section we briefly review a class of inflationary models where
the period of accelerating cosmological expansion is described by a
single dynamical parameter, the inflation field (see e.g. \cite{LiLy00}) and derive an expression for the spectrum of primordial gravitational waves as a function of the model parameters.  
Such analysis can be generalised to multi-field inflationary models, cf.
e.g.~\cite{Lid97}. Throughout the paper we adopt geometrical units in which 
$c = G = 1$.

The dynamics of a homogeneous and isotropic scalar field $\phi$ in a
cosmological background described by the Friedmann-Robertson-Walker
metric is determined by the equation of motion

\begin{equation}
\ddot{\phi} + 3H \, \dot{\phi} + V^{\prime}(\phi) = 0 \,,
\end{equation}
where $a$ is the scale factor, $H = \dot{a} / a$ the expansion rate and
$V(\phi)$ the scalar field potential; in the previous equation dots
refer to time derivatives and primes to derivatives with respect to
$\phi$. The evolution of $a$ is encoded into the Friedmann equation,

\begin{equation}
H^2 = \frac{8\pi}{3 \, m^2_{pl}} \left[ \frac{\dot{\phi}^2}{2} + V(\phi)
\right] \,,
\end{equation}
where $m_{pl} \sim 10^{19}$~GeV is the Planck mass. Inflation is a
period of accelerated expansion where $\ddot{a} / a \, > \, 0$ which
implies that the \textit{slow-roll parameters},

\begin{eqnarray}
\epsilon & = & \frac{m^2_{pl}}{16 \pi} \left( \frac{V^{\prime}}{V}
\right)^2\,,
\label{epsilon}\\
\eta & = & \frac{m^2_{pl}}{8 \pi} \left( \frac{V^{\prime\prime}}{V}
\right)\,,
\label{eta}
\end{eqnarray}
must be less than 1. 

Inflation generates two types of metric perturbations: (i) \textsl{scalar
or curvature perturbations}, coupled to the energy momentum tensor of
the matter fields, that constitute the seeds for structure formation and
for the observed anisotropy of the CMB and (ii) \textsl{tensor or
gravitational wave perturbations} that, at first order, do not couple
with the matter fields. Tensor perturbations are responsible for a CGWB.
In the slow-roll regime ($\epsilon \,, \eta < 1$), the power spectra of
curvature and tensor perturbations are given by
\begin{eqnarray}
\Delta^2_{{\cal R}} & = & \left[ \frac{H}{\dot{\phi}} \left(
\frac{H}{2\pi} \right) \right]^2_{k=aH} \,,
\label{Deltar}\\
\Delta^2_T & = & \frac{16}{\pi} \left( \frac{H}{m_{pl}} \right)^2_{k=aH} \,,
\label{Deltat}
\end{eqnarray}
where $\Delta^2_{{\cal R}}$ and $\Delta^2_T$ are functions of the
comoving wavenumber $k$ evaluated when a given mode crosses the causal
horizon $(k = aH)$. The spectral slopes of the scalar and tensor
perturbations are then given by
\begin{eqnarray}
n_s -1 & = & \frac{\rmd \ln \Delta_{{\cal R}}^2} {\rmd \ln \, k} \,,
\label{ns}\\
n_T & = & \frac{\rmd \ln \Delta_{T}^2} {\rmd \ln \, k} \,;
\label{nt}
\end{eqnarray}
$n_s$ and $n_T$ can also be written in terms of the slow-roll parameters
$\epsilon$ and $\eta$ as
\begin{eqnarray}
n_s & = & 1 - 6 \epsilon + 2 \eta \,,
\label{newns}\\
n_T & = & -2 \epsilon \,.
\label{newnt}
\end{eqnarray}
For single field slow-roll inflationary models the full set of metric
perturbations is described in terms of the quantities $\Delta_{{\cal
R}}$, $\Delta_T$, $n_s$ and $n_T$, which are however not independent.
Using Equations~(\ref{epsilon})-(\ref{Deltat}),(\ref{newns})
and~(\ref{newnt}) one finds the consistency relation
\begin{equation}
n_T = -\frac{r}{8} \,,
\label{consist}
\end{equation}
where 
\begin{equation}
r = \frac{\Delta^2_T}{\Delta^2_R}
\label{ratio}
\end{equation}
is the so-called tensor-to-scalar ratio. 

The spectrum of a cosmological
gravitational wave stochastic background is defined as
\begin{equation}
\Omega_\mathrm{gw}(f) = \frac{1}{\rho_c} \frac{\rmd
\rho_\mathrm{gw}}{\rmd \ln f} \,,
\end{equation}
where $\rho_\mathrm{gw}$ is the gravitational waves energy density,
$f = k / 2 \pi$ is the physical frequency and $\rho_c = 3 H^2_0 / 8 \pi$
is the critical energy density today. $H_0$ is the Hubble parameter and
$h_0 \equiv H_0/100 \, \mathrm{km} \, \mathrm{sec}^{-1} \,
\mathrm{Mpc}^{-1}$, so that $h_0^2 \Omega_\mathrm{gw}(f)$ is
independent of the value of the Hubble constant.

For the class of single-field, slow-roll inflationary models considered
here, the spectrum of a CGWB is given by~\cite{Lid97}
\begin{equation}
\Omega_\mathrm{gw}(f) = \frac{1}{24} \Delta^2_T \frac{1}{z_\mathrm{eq}}
\left( \frac{f}{f_0} \right)^{n_T}
\label{spectrum}
\end{equation}
where $z_\mathrm{eq} \approx 2.4 \times 10^4$ is the redshift of
matter-radiation equality and $f_0$ a reference frequency. In order to
be consistent with the recent analysis carried out by the WMAP team, in
this paper we choose $f_0 = 3.1 \times 10^{-17}$~Hz, corresponding
to a wavenumber $k_0 = 0.002 \; \mbox{Mpc}^{-1}$. Using the
Equations~(\ref{consist}) and~(\ref{ratio}), the
spectrum~(\ref{spectrum}) can be written as~\cite{Turner97}
\begin{equation}
\Omega_\mathrm{gw}(f) = \Omega_0 \, r \, A \exp \left[ {\cal N}(f) \,
n_\mathrm{gw}(f) \right] \,,
\label{newspectrum}
\end{equation}
where 
\begin{eqnarray}
{\cal N} & \simeq & 28.8 + \ln \left( \frac{f}{10^{-4} \mbox{Hz}}
\right) \,, \\
n_\mathrm{gw} & = & -\frac{r}{8} \left\{ 1 + \frac{\mathcal{N}}{2} \,
\left[ (n_s-1) + \frac{r}{8} \right] \right\} \,,
\end{eqnarray}
and $\Omega_0 = 5.1 \times 10^{-15}$. In Equation~(\ref{newspectrum}) the parameter $A$
accounts for the power spectrum normalisation with respect to the COBE
results: this parameter is currently constrained by the
measurements of CMB anisotropy to $A \sim 0.7 - 1.1$~\cite{Sper03}. Moreover,
since the GW spectrum is extrapolated over a wide range of scales, in
Equation~(\ref{newspectrum}) we have included the first order correction
for the running of the tensor spectral slope. Notice that
Equation~(\ref{newspectrum}) is valid provided that 
\begin{equation}
\left| (n_s-1) + \frac{r}{8} \right| \, \ll \, \frac{2}{\mathrm{max} \,
\cal{N}} \,.
\label{constraint}
\end{equation}
For $n_s = 1$ and $r \ll 1$, Equation~(\ref{newspectrum}) gives
$\Omega_\mathrm{gw}(f) \approx 3.7 \times 10^{-17} \, (r/10^{-2})$,
where we have set $A = 0.7$. For single-field inflationary models
$\Omega_\mathrm{gw}(f)$ is therefore described by two ``primordial
parameters'', $n_s$ and $r$, and one parameter $A$ which encodes the
effects due to the late cosmological evolution, such as the nature of
the dark energy component. In this paper we set $A = 0.7$ and consider
the gravitational wave spectrum $\Omega_\mathrm{gw}(f)$ as described by
two unknown parameters, $n_s$ and $r$, that need to be determined by
observations.

\section{Testing inflationary models with the Big-Bang-Observer}
\label{sec:results}

The Big-Bang-Observer is presently envisaged as a constellation of four
3-arm space-based interferometers on the same Heliocentric orbit at the
vertices of an equilateral triangle, with two interferometers co-located
and rotated by $180^\circ$ at one of the vertices. The arm length of the
interferometers is about $5 \times 10^{4}$ km (a hundredth of the LISA
arm length) corresponding to a peak sensitivity at $\sim 1$~Hz.
Different parameters have been suggested for the instrument, which in
turn correspond to different sensitivities; following~\cite{bbo} we
consider three possible choices, that we summarise in
Table~\ref{tab:bbo-param}; we call the corresponding mission concept as
``BBO-lite'', ``BBO-standard'' and ``BBO-grand''. In this Section we explore the
region of the parameter space $(n_s, r)$ that can be probed with an
instrument of the BBO class and how it depends on the instrumental
parameters; we also compare the sensitivity of a gravitational wave
mission with the information that can be obtained indirectly from CMB
observations using WMAP, Planck~\cite{planck} and CMBPol~\cite{cmbpol}.

Gravitational wave searches for stochastic backgrounds are optimally
carried out by cross-correlating the data sets recorded at different
instruments, which allows to disentangle the common stochastic
contribution of a CGWB from the (supposedly uncorrelated) contribution
from the instrumental noise~\cite{AR99}. The signal-to-noise ratio can be
efficiently built only when the separation of two instruments is smaller
than (half of) the typical wavelength of the waves (in the BBO case $\lambda
\approx 10^{11}$~cm), and therefore only the co-located instruments
can be used in the BBO mission to carry out highly sensitive searches of
stochastic signals. The other interferometers of the constellation allow
to accurately identify individual sources and subtract any contaminating
radiation from the data streams. Assuming that the noise of the
instruments is uncorrelated, stationary and Gaussian, the optimal
signal-to-noise ratio $\mathrm{S/N}$ that can be achieved
is~\cite{UV01}

\begin{eqnarray}
\mathrm{S/N} & \approx & \frac{3 H_0^2}{10 \pi^2} \sqrt{T} \left[
\int_{-\infty}^{\infty} \rmd f \frac{\gamma^2(f)
\Omega_{\mathrm{gw}}^2(f)}{f^6 S_h^{(1)}(f)\, S_h^{(2)}(f)}
\right]^{1/2} \,,
\nonumber\\
& \approx & 3 \, \left( \frac{h_0^2\Omega_\mathrm{gw}}{10^{-15}}
\right) \, \left[ \left( \frac{\Delta f}{1 \, \mathrm{Hz}} \right) \,
\left( \frac{T}{10^8 \, \mathrm{s}} \right) \, \right]^{1/2} \left(
\frac{f}{1 \, \mathrm{Hz}} \right)^{-3} \, \left( \frac{S_h}{10^{-48} \,
\mathrm{Hz}^{-1}} \right)^{-1} \,,
\label{eqn:snr}
\end{eqnarray}
where $S_h^{(1, 2)}$ is the power spectral density of the detectors noise
-- in the remaining of the paper we assume the instruments to have
identical sensitivity and therefore set $S_h^{(1)}(f) = S_h^{(2)}(f) =
S_h(f)$ -- $T$ is the integration time, $\Delta f$ is the effective
bandwidth over which the signal-to-noise ratio is accumulated and
$\gamma(f)$ is the overlap reduction function~\cite{UV01}. In
Table~\ref{tab:bbo-param} we report the frequency at which the noise of
BBO reaches the minimum and the corresponding value of $S_h$, depending
on the choice of the instrumental parameters.

\begin{table}
\caption{\label{tab:bbo-param} Possible instrumental parameters of the
proposed Big-Bang-Observer mission~\cite{bbo}: laser power $P_{La}$ and
wavelength $\lambda$, optical efficiency $\epsilon$, mirror diameter
$D$ and the ratio of the BBO acceleration noise to that of LISA $\eta$.
Using these parameters it is straightforward to derive the noise spectral
density $S_h(f)$ from~\cite{generator}: accordingly we report the
frequency $f_\mathrm{*}$ at which the noise reaches its minimum and the
relevant value $S_\mathrm{*} = S_h (f_\mathrm{*})$.}
\begin{indented}
\lineup
\item[]\begin{tabular}{l|cccccccc}
\br                              
Configuration & $P_{La}$ & $\lambda$ & $\epsilon$ & L & $D$ & $\eta$ &
$f_\mathrm{*}$ & $S_\mathrm{*}^{1/2}$\\
& (W) & ($\mu$~m) & & (km) & (m) & & Hz & $\mathrm{Hz}^{-1/2}$\\
\mr
BBO-lite & 100 & 1.06 & 0.3 & $2 \times 10^4$ & 3 & 0.1 & 1.3 & $5.5
\times 10^{-24}$\\
BBO-standard & 300 & 0.5 & 0.3 & $5 \times 10^4$ & 3.5 & 0.01 & 0.6& $7.9 \times
10^{-25}$\\
BBO-grand & 500 & 0.5 & 0.5 & $2 \times 10^4$ & 4 & 0.001 & 0.7 & $3.3
\times 10^{-25}$\\
\br
\end{tabular}
\end{indented}
\end{table}

\begin{figure}[htbp]
\vspace{3pt}
\begin{center}
 \includegraphics[height=12cm]{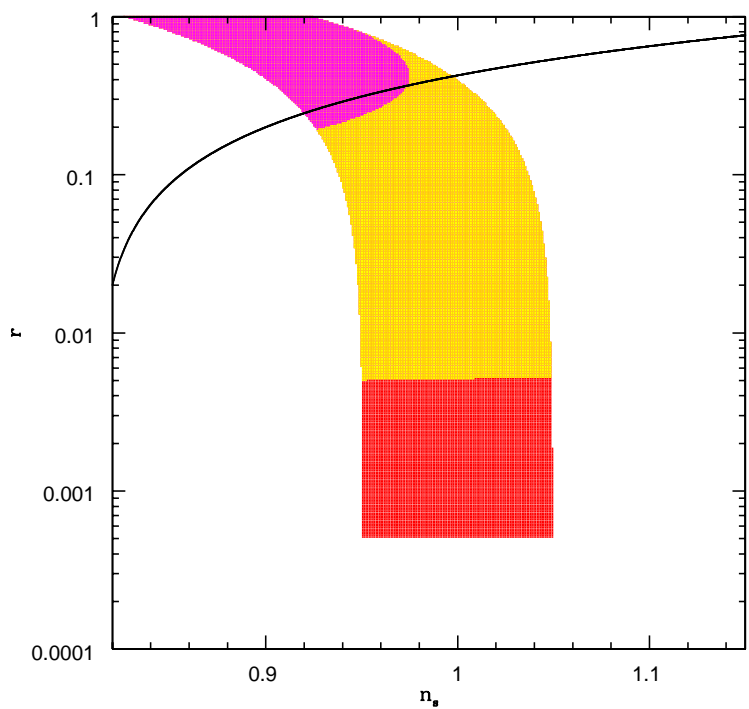}
\end{center}
\caption{The sensitivity of the Big-Bang-Observer mission to a cosmic
gravitational wave background generated by a single field slow-roll
inflationary model. The plot shows the region in the parameter space
$r$ and $n_s$ (see Section 2) that can be detected (corresponding to a
false alarm alarm probability of $1\%$ and false dismissal rate of
$10\%$). The magenta, yellow and red  regions correspond to the limits
obtained in a 3 yr long observation with BBO-lite, BBO-standard, and BBO-grand,
respectively. The solid line correspond to the present best upperlimit
set by WMAP observations~\protect{\cite{Kinneyetal04}}}
\label{fig:snr}
\end{figure}

We have computed the signal-to-noise ratio, Equation~(\ref{eqn:snr}),
generated by a single-field inflationary spectrum
$\Omega_{\mathrm{gw}}(f; n_s, r)$, Equation~(\ref{spectrum}) for the
three BBO configurations reported in Table~\ref{tab:bbo-param}. The
parameters of the signal model have been chosen in the range $1.2 \le
n_s \le 0.8$ and $0\le r \le 1$ and satisfy the constraint given by
Eq.~(\ref{constraint}). We have assumed an effective integration
time of 3 years and the noise spectral density has been derived using
the Sensitivity Curve Generator for Space-borne Gravitational Wave
Observatories~\cite{generator} with the parameters reported in
Table~\ref{tab:bbo-param}. Figure~\ref{fig:snr} summarises the results
and compare them with the current upper-limits on $n_s$ and $r$ which
have been inferred from the analysis of the WMAP
data~\cite{Kinneyetal04}. The first interesting result is that the
BBO-lite configuration would not be able to improve our understanding of
standard inflation beyond what is already known; in fact the sensitivity
of BBO-lite is broadly comparable to the limit currently set by WMAP.
This has an immediate implication on the technology programme that will
lead to a BBO-like mission: the parameters reported in
Table~\ref{tab:bbo-param} for BBO-lite are simply too conservative and
would not allow us to achieve the mission science goal.

On the other hand the BBO-standard configuration is able to probe the
entire range of $n_s$ and to reach values of the scalar-to-tensor ratio
$r \approx 5 \times 10^{-3}$ for a 1\% false alarm and 10\% false
dismissal rate; by adopting the BBO-grand configuration it would be
possible to do even better and reach $r \approx 5 \times 10^{-4}$. Notice
that for $r \ll 1$, the minimum value of the scalar-to-tensor ratio
$r_\mathrm{min}$ that can be observed scales as $r_\mathrm{min} \sim
1/S_h$, every other parameter being equal. Not surprisingly a dedicated
mission such as BBO would improve our ability of probing the range of
unknown parameters by (roughly) three orders of magnitudes, with respect to
current limits. However, CMB experiments such as Planck (2007) and, in the 
more distant future, CMBPol will also be in a position of searching for the
signature of a CGWB and it is worth comparing the sensitivity that can
be achieved by means of indirect observations with the BBO results.

\begin{figure}
\begin{center}
\mbox{
%\resizebox{10 cm}{7 cm}
\scalebox{0.4}{\rotatebox{360}{\includegraphics{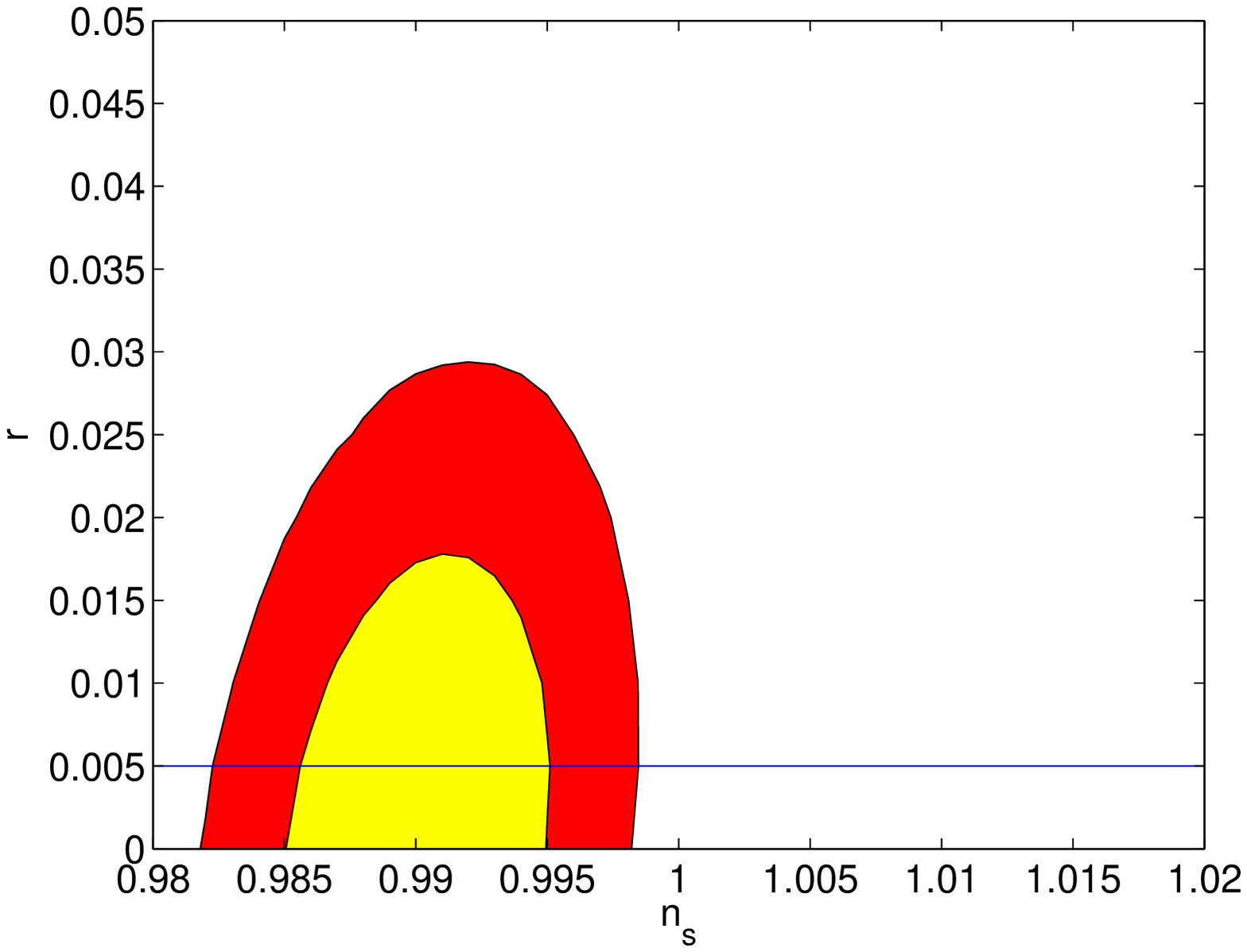}}}
\scalebox{0.4}{\rotatebox{360}{\includegraphics{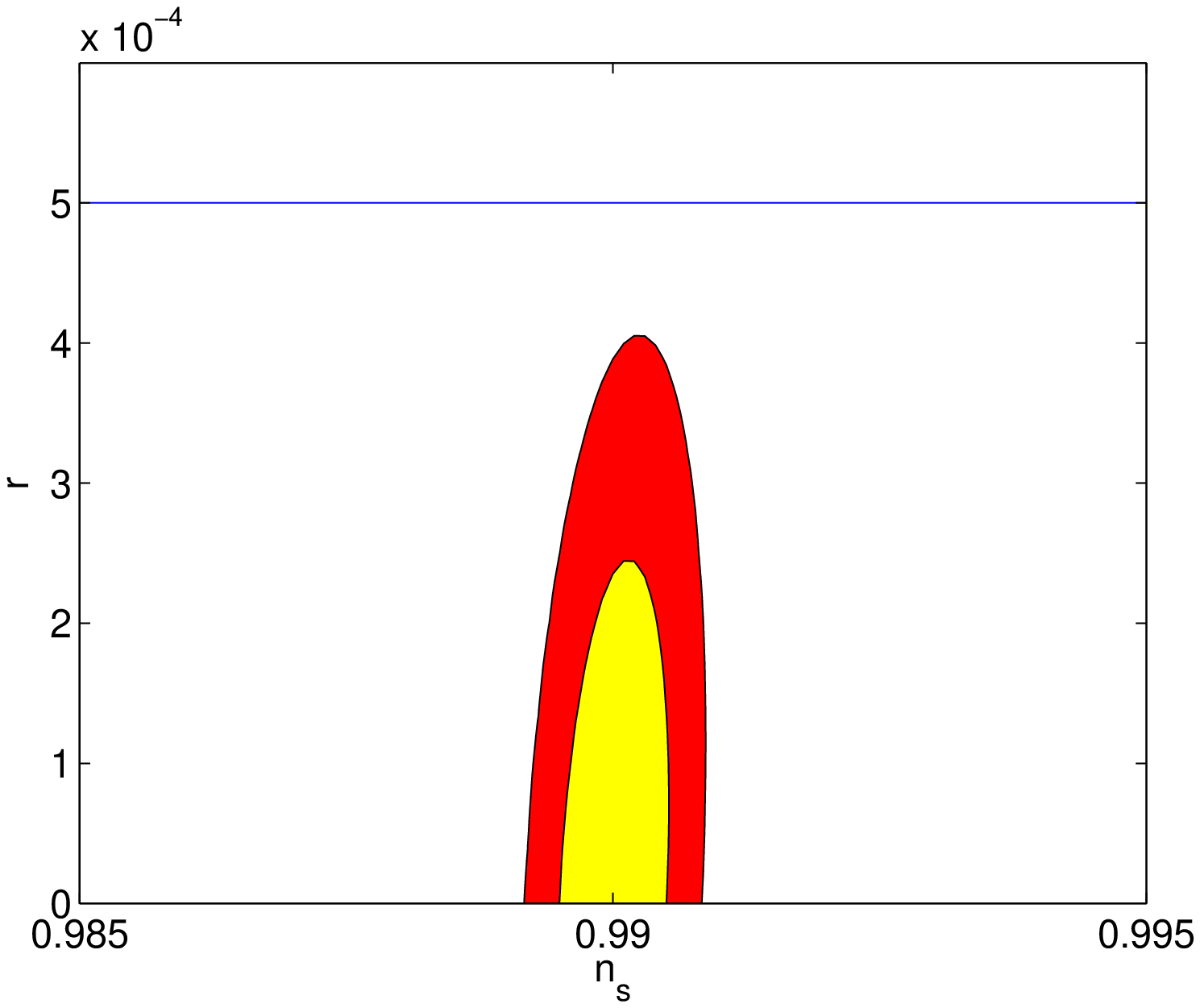}}}
}
\end{center}
\caption{\label{fig:comp}The sensitivity of Planck and CMBPol to
indirect observations of a cosmic gravitational wave background produced
during inflation. The plots show the region of the the parameter space
$r$ and $n_s$ corresponding to the 68\% and 95\% confidence level
upper-limit to a CGWB (red and yellow areas, respectively). The left
plot corresponds to Planck observations and the line refers to the detection limit
obtained with the BBO-standard configuration, cf Figure~\ref{fig:snr}. The
plot on the right corresponds to CMBPol observations and the line
refers to the detection limit obtained with the BBO-grand configuration, cf
Figure~\ref{fig:snr}.}
\end{figure}

In order to make this comparison, we have determined the theoretical
confidence intervals on the parameters $n_s$ and $r$ by computing the
corresponding Fisher information matrix for Planck and CMBPol,
including both the polarisation and the temperature anisotropy CMB
spectra. More in detail, we have assumed as cosmology the best-fit
model consistent with WMAP data~\cite{Sper03} and we have
marginalised with respect the ionisation optical depth in order to take
into account its effect on  the B-mode polarisation. For Planck, we have
assumed an average pixel sensitivity of $11.6\, \mu$K and $24.3\, \mu$K 
for the temperature and polarisation anisotropies respectively, 
while for CMBPol the corresponding noise levels are reduced by a factor 40.
Figure~\ref{fig:comp} summarises the results: we show the regions in
the two-dimensional parameter space $(n_s, r)$ corresponding to the
68\% and 95\% confidence level for the null hypothesis (i.e.~no CGWB)
for Planck and CMBPol and compare it with the limit of BBO and
BBO-grand observations, respectively (those reported in
Figure~\ref{fig:snr}).

One important caveat is that the results that we have presented so far,
both for direct and indirect observations, are computed assuming that
the only factor limiting the sensitivity of the instruments is the
intrinsic noise of the detectors, whereas other effects could actually
provide the limitation. Astrophysical foregrounds and radiation from
individual GW sources can limit the sensitivity of BBO. Stochastic
foregrounds are produced by the incoherent superposition of radiation
from large populations of astrophysical sources. Foregrounds are
particularly dangerous, because they provide a \emph{fundamental}
sensitivity limit for the mission~\cite{UV01}. In the BBO band, the
strongest contributions, according to our present astrophysical
understanding come from rotating neutron stars and supernovae generated
by population III objects\cite{Buonannoetal04}. Foregrounds from rotating neutron stars
should not be a serious limitation, as their contribution to the
spectrum is $\Omega_\mathrm{gw} \simlt 10^{-22}$. On the other hand
supernovae from  population III objects could be a very serious obstacle to
achieve high sensitivity and might overwhelm the signal produced by
inflation. In fact they could produce a foreground with intensity
$h_0^2\Omega_\mathrm{gw} \sim 10^{-18}$ at $f \sim 1$~Hz. For comparison this
is equivalent to a CGWB with $r \sim 10^{-3}$. Even assuming that no
foreground is sufficiently strong to compete with the signal from
inflation, deterministic signals, primarily from binary neutron stars up to
high redshift, will be present in the data set and need to be identified
and removed to a high degree of precision in order not to introduce spurious effects.

On the other hand the sensitivity of CMB experiments to primordial
gravitational waves strongly depends on the distinctive signature
produced by a CGWB on the B-mode of the CMB polarisation. Indeed the
B-mode polarisation is a particular sensitive probe of primordial tensor
perturbations, since it does not receive contributions from primordial
density perturbations. However, gravitational lensing by cosmological
structure also generates a B-mode component in the CMB polarisation
~\cite{ZS98} and such foreground cannot be fully subtracted. The lensing
contamination poses a fundamental limit on the sensitivity to a B-mode
component due to primordial gravitational waves, corresponding to a
lower limit  on the scalar-to-tensor ratio $r$ of about $6 \times
10^{-4}$~\cite{KS02,KCK02}.

\section{Conclusions}
\label{sec:conclusions}

The direct detection of a cosmological gravitational wave stochastic
background produced during inflation is of great importance for the
understanding of early Universe cosmology and shall provide a direct
test of one of the fundamental, and not yet probed predictions of
inflationary theories. In this paper we have explored the sensitivity of
the Big-Bang-Observer mission to backgrounds generated by slow-roll,
single field inflationary models and compared it with indirect
limits that future CMB missions, such as Planck and CMBPol are
expected to set. Our analysis shows that mild technological improvements considered
for the BBO-lite configuration would not meet the science goals of a dedicated 
gravitational wave interferometric mission; on the other hand the ambitious
choices of the instrumental parameters for the standard and grand configuration
of BBO would allow us to achieve a sensitivity $h_0^2\Omega_\mathrm{gw} \sim 10^{-19}$
in the frequency band $0.1\,\mathrm{Hz} - 1\,\mathrm{Hz}$. This value is broadly 
comparable with what could be achieved by one of the inflationary probes for CMB 
observations, such as CMBPol that are currently being discussed. 

It is however important to stress that throughout this paper we have assumed
that the effect of foreground emission from unresolved sources and/or lensing would 
have a negligible impact on the sensitivity of the missions. This hypothesis is
useful to gain an insight into the ultimate performance of the experiments, but its
range of validity needs to be careful investigated. For direct gravitational wave
observations it is clear that at some point astrophysical foregrounds will provide the 
fundamental sensitivity limit. What is the level at which this can occur and the
consequences on our ability of testing prediction needs to be parametrised as a function
of our (still poor) knowledge of the relevant astrophysical scenarios.

\end{document}